\documentclass[prl, aps, twocolumn, superscriptaddress, showpacs]{revtex4}

\usepackage{slashed}
\usepackage{graphicx}
\usepackage{subfigure}
\usepackage[usenames, dvipsnames]{color}
\usepackage{graphics}
\usepackage{hyperref}
\usepackage{bm}
\usepackage{amsmath}
\usepackage{color}
\usepackage{amsfonts}
\hypersetup{backref,
colorlinks=true,
linkcolor=cyan,
linktoc=page,
citecolor=cyan,
urlcolor=cyan}

\everymath{\displaystyle}

\begin{document}

\title{ Violation of the $\cal  PT-$symmetry  and structure formation in the dark matter $-$gravitational wave interaction }
\author{Catarina Bastos}
\email{catarina.bastos@tecnico.ulisboa.pt}
\affiliation{GoLP/Instituto de Plasmas e Fus\~ao Nuclear, Lisbon, Portugal}

\author{Hugo Ter\c cas}
\email{hugo.tercas@it.lx.pt}
\affiliation{Instituto de Plasmas e Fus\~ao Nuclear, Lisbon, Portugal}
\affiliation{Instituto de Telecomunica\c{c}\~oes, Lisbon, Portugal}


\begin{abstract}

In flat spacetime, quantum fluctuations in dark matter, as described as a Bose-Einstein condensate, are stable and display a relativistic Bogoliubov dispersion relation. In the weak gravitational field limit, both relativistic and nonrelativistic models self-gravitating dark matter suggest the formation of structures as the result of a dynamical (Jeans) instability. Here, we show that in the presence of spontaneous symmetry breaking of the dark matter field, the gravitational wave is damped for wave-lengths larger than the Jeans length. Such energy is converted to the Bogoliubov modes of the BEC that in their turn become unstable and grow, leading to the formation of structures even in the absence of expansion. Remarkably, this compensated attenuation/amplification mechanism is the signature of a discrete $\cal PT$ symmetry-breaking of the system. 

\end{abstract}
\maketitle

{\it Introduction.}
Scalar field theories in curved spacetimes are on the basis of modern advances in cosmology and astrophysics \cite{kolb_1989, zee_2003, dodelson_2003}, as they constitute important candidates to explain the behaviour of dark-matter \cite{chavanis_2011, harko_2011a, suarez_2014, chavanis_2015a}. Their relativistic dynamics is governed by the Klein-Gordon (KG) equation \cite{klein_gordon}, historically emerging as a first attempt to unify quantum mechanics with the special relativity theory, to obtain a unified theory to explain our universe.
The Klein-Gordon-Einstein (KGE) equations, that involve a coupling between the KG equations and gravity, were first used to study boson stars \cite{bosonstars}. Although, the first models of these stars do not assumed that bosons could have a self interaction potential, it was shown then that self-interaction can significantly change the physical dimensions of the boson stars and make them clearly more interesting as an astrophysical object \cite{colpi_1986}. Furthermore, models of dark matter (DM) halos were proposed based on scalar fields that are described by KGE equations \cite{suarez_2014}. These DM halos can be explained through Schr\"odinger-Poisson (SP) equation or Gross-Pitaevskii-Poisson (GPP) equations, since  the Newtonian limit is valid at the galactic scale. In this sense, we can think about DM halos as gigantic quantum objects made of Bose-Einstein condensates (BECs). Furthermore, these models are a tentative of solving the problem of Cold Dark Matter (CDM), as the wave properties of bosons can stabilize the system against gravitational collapse.
At the cosmological level it is quite important to study the implications of these scalar field models. It was shown by Matos et al. \cite{matos_2009} that when a spatially homogenous interacting real scalar field competes with baryonic matter, radiation and dark energy in terms of cosmological evolution, these real scalar fields can reproduce quite well the cosmological predictions of the $\Lambda$-CDM model. A perturbative analysis then showed the formation of structures corresponding to DM halos. Finally, Chavanis \cite{chavanis_2011} has considered the case of a complex self-interacting scalar field in the context of Newtonian cosmology and based on the GPP equations. The formation of structures has been recently studied through the Jeans instability of an homogeneous self-gravitating BEC in a static background \cite{chavanis_2015}. Basically, the so-called BECDM have shown that perturbations grow faster than in a $\Lambda$-CDM model. Some relativistic models have been then analyzed \cite{chavanis_2011, harko_2011b}.

In the last year, gravitational waves (i.e. fluctuations in the metric) generated by accelerated mass distributions, like massive black holes, were finally detected by the LIGO collaboration \cite{LIGO2015}. The theory of general relativity predicts that the amplitude of these gravitational is extremely small, which harnessed their detection for a long time. Although gravitational waves produced by black hole collisions could be detected, finding experimental evidence of primordial gravitational waves remains elusive. Nevertheless, the advent of table-top, high-sensitivity devices based on quantum technologies revived their interest. Also, ESA is now developing the eLiSA, a space-based interferometer that will be used to detect gravitational waves in other range of frequencies, the low-frequency band \cite{amaro-seoane2012}. It is expected that it could detect waves coming from other sources rather than merging black holes.
On the other hand, a lot of attention has been drawn to the study of space-time effects in quantum systems as, for instance, in phononic fields \cite{ahmadi2013, sabin2014}. It was shown in that gravitational waves can create phonons in a BEC \cite{sabin2014}, a features that is motivating a new generation of gravitational-wave detectors using matter waves, which may become a reality in a medium-term timescale. 

In this work, we investigate the self-consistent dynamics of an interacting complex scalar field (BEC) - described by a nonlinear Klein-Gordon equation - evolving in a fluctuating space-time. In particular, we show that gravitational waves (obtained from Einstein's equation in the weak field limit) can couple to scalar field fluctuations, leading the dynamics of the latter unstable. Remarkably, the presente instability mechanism appears to be associated with the violation of the discrete parity-time ($\mathcal{PT}$) symmetry. The latter, initially proposed as a concept in quantum mechanics \cite{bender_98}, is now being extensively studied in optics \cite{zyablovsky_2014, suchkov_2015, konotop_2016} and, more recently, in acoustics \cite{zhu_2014}. In such systems, any spatial region with a loss is mirrored by a region of gain. Therefore, the processes of light (or sound) absorption and amplification can be compensated, and the frequencies of the eigen optical (acoustic) modes can be real. When $\mathcal{PT}$-symmetry is broken, the eigenmodes appear in complex conjugate pairs. In our case, complex eigenmodes appear for hybrid modes made of the mixture between gravity and Bogoliubov (sound) BEC modes. This suggests that the formation of structures, as described by the Jeans self-gravitating instability, due to primordial gravitational waves is a consequence of the breaking of the $U(1)\times \mathcal{PT}$ symmetry. Our findings show that the gravitational wave is damped for wave lengths larger than the Jeans length and the energy is converted to the Bogoliubov modes of the BEC, which grow in time. This will turn the system unstable, leading to the formation of primordial cosmological structures  even in the absence of an expanding universe. Moreover, we argue that this particular form of the space-time$-$field interaction may be an important mechanism preventing the detection of primordial gravitational waves, as their energy is transferred to the matter field originating structures in the universe. 

{\it Minimally coupled theory.}
The dynamics of a complex scalar field (SF) $\varphi(x_\mu)$ in a curved spacetime of curvature $R$ is governed by the following minimal-coupling action
\begin{equation}
\mathcal{S}=\frac{c^4}{16\pi G}\int d^4x \sqrt{-g} R+\int d^4 x\sqrt{-g} \mathcal{L_\varphi},
\label{eq_action}
\end{equation}
where $x_\mu=(-c t, \textbf{x})$ is the four vector, $g_{\mu \nu}$ is the metric tensor, $g=g^\mu_\mu$ denotes its trace, and $\mathcal{L}_\varphi$ is the SF Lagrangian 
\begin{equation}
\mathcal{L}_\varphi=\frac{1}{2}g^{\mu\nu}\partial_\mu \varphi^*\partial_\nu \varphi-V\left(\vert \varphi\vert^2\right).
\end{equation}
Here, $V\left(\vert \varphi\vert^2\right)$ contains the KG rest mass term and the self-interaction potential,
\begin{equation}
V\left(\vert \varphi\vert^2\right)=\frac{m^2 c^2}{2\hbar^2}\vert \varphi\vert^2+\frac{1}{4}\lambda \vert \varphi\vert^4,
\end{equation}
where $\lambda=8 \pi  a_s m/\hbar^2$ is the coupling constant, $a_s$ is the scattering length and $m$ is the field mass. The minimization of Eq. \eqref{eq_action} with respect to $\varphi$ provides the Euler-Lagrange equation 
$\nabla_\mu \left[ \frac{\partial \mathcal{L}_\varphi}{\partial \left( \partial_\mu \varphi\right)^*}-\frac{\partial \mathcal{L}_\varphi}{\partial\varphi^*}\right]=0$, which in turn yields the following generalized KG equation
\begin{equation}
\square_g \varphi+V'\left(\vert \varphi \vert^2\right)_{,\varphi^*}=0,
\label{eq_KG1}
\end{equation}
where $\square_g\equiv g^{\mu\nu}\partial_\mu \partial_\nu-g^{\mu\nu}\Gamma_{\mu\nu}^\alpha \partial_\alpha$ is the generalized d'Alembert operator and $\Gamma_{\mu \nu}^\alpha$ denotes the Christoffel symbol \cite{schutz}. In Eq. \eqref{eq_KG1}, we made use of the parallel transport of the metric $\nabla_\mu g^{\mu \nu}=0$. A similar minimization procedure with respect to the metric $g_{\mu \nu}$ leads to the Einstein field equations
\begin{equation}
R_{\mu \nu}-\frac{1}{2}g_{\mu\nu} R=\kappa T_{\mu\nu}, 
\label{eq_E1}
\end{equation}
with $\kappa= 8\pi G/c^4$ and $T_{\mu\nu}$ is the energy-momentum tensor
\begin{equation}
\begin{array}{ll}
T_{\mu \nu}=\frac{1}{2} &\left( \partial_\mu \varphi^* \partial_\nu \varphi +\partial_\nu \varphi^* \partial_\mu \varphi\right)\\
&-g_{\mu\nu}\left[ \frac{1}{2} g^{\alpha \beta}\partial_\alpha \varphi^*\partial_\beta \varphi-V\left(\vert \varphi\vert^2\right)\right].
\end{array}
\end{equation}

{\it Perturbative analysis.}
We now assume a perturbation around the Minkowski space time of the form $g_{\mu\nu}=\eta_{\mu\nu}+h_{\mu\nu}$, where $h_{\mu\nu}\ll \eta_{\mu\nu}$ is the spacetime ripple and $\eta_{\mu\nu}={\rm diag}(-,+,+,+)$. To first order in $h_{\mu\nu}$, Eq. \eqref{eq_KG1} reads
\begin{equation}
\square \varphi +V'\left(\vert \varphi \vert^2\right)_{,\varphi^*}+h^{\mu\nu}\partial_\mu\partial_\nu\varphi -\eta^{\mu \nu}\gamma_{\mu\nu}^\alpha \partial_\alpha \varphi=0,
\label{eq_KG2}
\end{equation}
where $\gamma_{\mu\nu}^\alpha=\frac{1}{2}\left( \partial^\alpha h_{\mu\nu}+\partial_\nu h_{\mu}^{\alpha}-\partial_\mu h_{\nu}^{\alpha}\right)$, with $h_{\mu}^{\nu}=\eta^{\alpha \nu}h_{\mu\alpha}$. Making use of the trasnverse-traceless (TT) gauge, $\partial_\mu h_{\nu}^{\mu}=0$ and $h\equiv h_{\mu}^{\mu}=0$, the last term in Eq. \eqref{eq_KG2} vanishes and the KG equation explicitly reads
\begin{equation}
\square \varphi+\frac{m^2 c^2}{\hbar ^2}\varphi+\lambda \vert \varphi\vert^2 \varphi+h^{\mu\nu}\partial_{\mu}\partial_{\nu}\varphi=0,
\label{eq_KG3}
\end{equation}
where we made use of the property $h^{\mu\nu}=h_{\mu\nu}$. Similarly, the weak-field limit of Eq. \eqref{eq_E1} describes spacetime radiation (gravitational waves) in the presence of matter
\begin{equation}
\square h_{\mu\nu}=-2\kappa T_{\mu\nu},
\label{eq_E2}
\end{equation}
with $\square=-(1/c^2)\partial_t^2+\nabla^2$. In what follows, we introduce quantum fluctuations around the homogeneous scalar field (i.e. the vacuum expectation value $\langle \varphi \rangle =\sqrt{n_0}$ spontaneously breaking the continuous $U(1)$ symmetry) in the form
\begin{equation}
\varphi(x_\mu)=\sqrt{n_0} e^{-i\mu t/\hbar}\left[ 1+\sum _k\left( u_k e^{ik_\mu x^\mu}+v_k^* e^{-ik_\mu x^\mu}\right)\right],
\label{eq_phi}
\end{equation}
where $\mu$ is the chemical potential of the condensate. By dividing the metric fluctuation into its time and space components, $h_{\mu,\nu}=h_{00}+h_{ij}\equiv 2\phi/c^2+h_{ij}$, we can obtain purely transverse solutions satisfying the condition $k^jh_{ij}=0$ as 
\begin{equation}
\left(\square -\frac{m_{\rm eff}^2 c^2}{\hbar^2}\right) h_{ij}=0,
\end{equation}
where $m_{\rm eff}^2=2\hbar ^2\kappa V(n_0)/c^2$ is the square of the effective graviton mass. Assuming plane-wave solutions of the form $h_{ij}=\sum_k \chi_{ij}e^{ik_\mu x^\mu}$, we obtain the dispersion relation
\begin{equation}
\omega^2 =\omega_p^2+c^2k^2,\quad \omega_p^2=\frac{m_{\rm eff }^2 c^4}{\hbar^2}.
\label{eq_plasma}
\end{equation}
The latter is obtained by making use of the equation of state obtained at the zeroth order, which fixes the chemical potential of the BEC as
\begin{equation}
\mu=\sqrt{m^2 c^4+\hbar^2 \lambda n_0 c^2}=mc^2\sqrt{1+\frac{c_s^2}{c^2}},
\end{equation}
where $c_s=\hbar /(m\sqrt{\lambda n_0 })$ is the BEC sound speed. The dispersion relation in Eq. \eqref{eq_plasma} is analogous of that of an electromagnetic wave propagating in a charged medium characterized by a plasma frequency $\omega_p$, where the photon also acquires an effective mass. Consequently, the KG equation decouples from Einstein's equations and the Bogoliubov modes for the scalar field can be obtained from Eqs. \eqref{eq_KG3} and \eqref{eq_phi}. Their dispersion relation can be found from plugging Eq. \eqref{eq_phi} in Eq. \eqref{eq_KG3} and separating it into its particle (anti-particle) coefficients $\propto e^{ik_\mu x^\mu}$ ($\propto e^{-ik_\mu x^\mu}$). The resulting secular equation contains two real solutions $\omega_{\pm}$ (and the corresponding ``anti-mode" solutions $\omega^*_{\pm}=-\omega_{\pm}$) 
\begin{equation}
\omega_\pm^2=2\omega_0^2+c^2 k^2 \pm 2\sqrt{\omega_0^4+c^2k^2\omega_0^2},
\label{eq_modes}
\end{equation}
where $\omega_0=mc^2\left(1+\beta^2\right)/\hbar$, with $\beta^2=3 c_s^2/(2c^2)$, is the cut-off frequency. In the long wavelength limit $k\ll \xi (1+\beta^2)$ - with $\xi=\hbar/m c_s$ denoting the healing length - the lower (Goldstone) mode is {\it gapeless}, $\omega_-\simeq c_s(1+\beta^2)k+\xi c_s k^2/2(1+\beta^2)^3$, reducing to the usual Bogoliubov dispersion in the non-relativistic limit $\beta \rightarrow 0$ \cite{pitaevskii_book, pethick_book}. The {\it gapped} mode, corresponding to the massive Higgs mode of mass $M=\hbar \omega_0/c^2=m(1+\beta^2)$, reads $\omega_+\simeq \sqrt{M^2 c_s^4/\hbar^2 +c_s^2 k^2}$. Although with a different notation, the dispersion modes of Eq. \eqref{eq_modes} have first been discussed in Ref. \cite{fagnocchi}. \par
\begin{figure}[t!]
\includegraphics[scale=0.59]{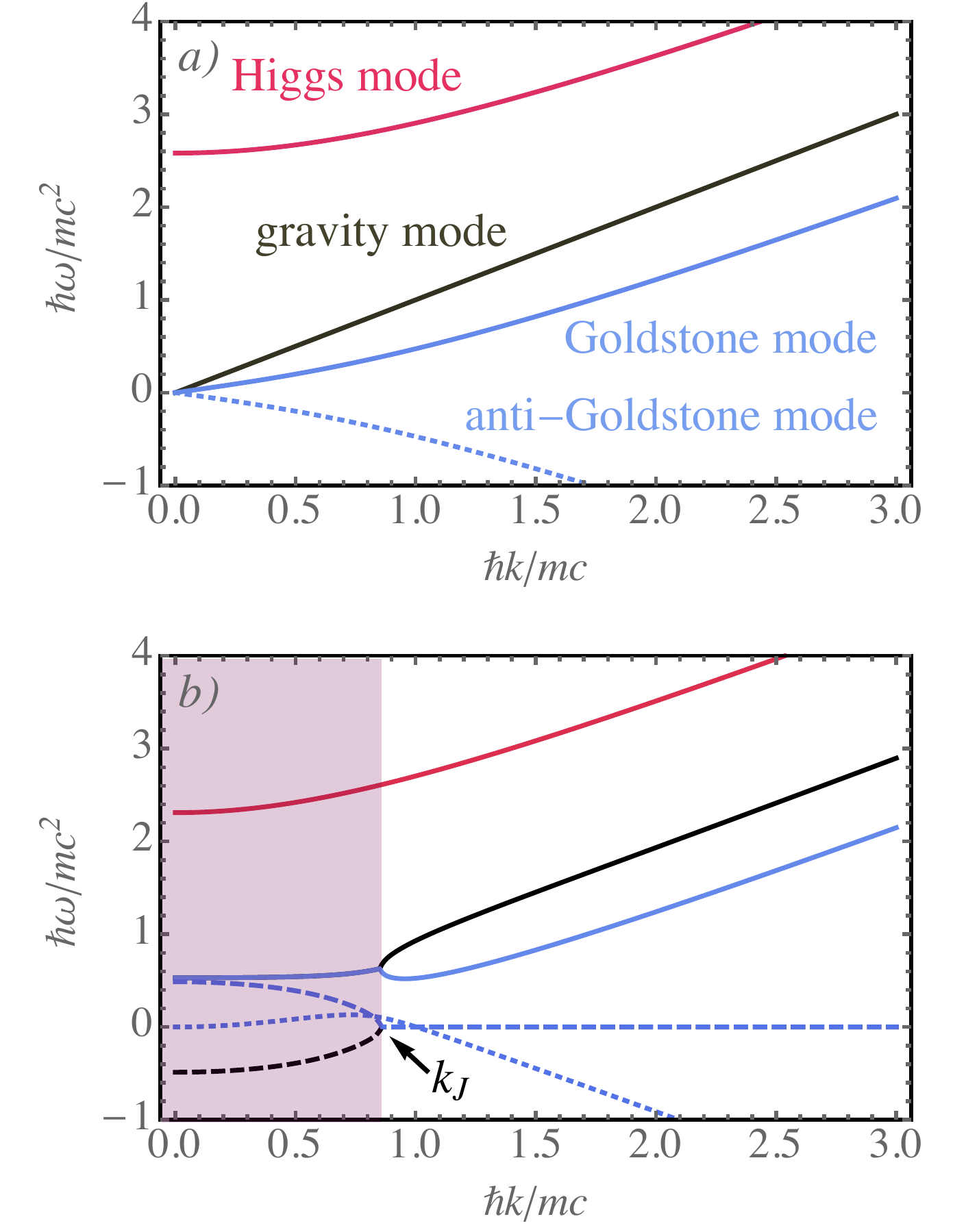}
\caption{Dispersion relation of the various modes present in the dynamics. Top panel: mode dispersion in the absence of coupling ($G=0$). We observe that the Bogoliubov-Goldstone modes are $\mathcal{PT}$-symmetric. Bottom panel: when the gravity is switched on, the Goldstone modes lose their symmetry. The gravity mode is damped to favour unstable (growing) modes in the BEC. The imaginary part of the frequencies goes to zero at the Jeans mode $k_J$, exhibiting the usual signature of $\mathcal{PT}-$ symmetry breaking. For illustration purposes, we use $c_s=c/3$.}
\label{fig_disp}
\end{figure}
The situation changes if we consider perturbations in the time-time components of Einstein's equations \eqref{eq_E1}, i.e., for gravitational waves of the form $h_{\mu\nu}\simeq h_{00}=2\phi/c^2$. This amounts to generalize the usual self-gravitating problem, as described by the Klein-Gordon-Poisson system \cite{chavanis_2015}, to the study of propagation of gravitational radiation in a symmetry-broken quantum vacuum. As we are about to see, the formation of structures emerges in this case as a consequence of the violation of the discrete $\mathcal{PT}-$ symmetry.  Putting Eqs. \eqref{eq_E2} and \eqref{eq_KG3} together, and keeping terms to the first order in the Fourier components of the vector $\bm{V}_k=(u_k, v_k, \phi_k)$, we obtain the eigenvalue problem $\mathcal{L}_k \bm{V}_k=0$, where
\begin{equation}
\mathcal{L}_k=\left[
\begin{array}{ccc}
\epsilon_1^2 &-m^2 c^2 c_s^2 &- 2\mu^2\\
-m^2 c^2 c_s^2 & \epsilon_2^2 & 0 \\
-\tilde\kappa \left(\mu -\hbar \omega \right)^2 &  -\tilde\kappa \left(\mu +\hbar \omega \right)^2 &\epsilon_0^2
\end{array}\right],
\end{equation}
with $\tilde \kappa= 8\pi G n_0/c^4$, $\epsilon_{1,2}^2=\left( \mu\pm \hbar \omega\right)^2-\hbar^2 c^2 k^2-m^2c^4-2m^2 c_s^4$, and $\epsilon_0^2=\hbar^2(\omega^2-c^2k^2)$. Nontrivial solutions are obtained by solving the secular equation $\det (\mathcal{L}_k)=0$ in respect to $\omega$, for which we obtain six solutions (three for positive-energy and other three for negative-energy excitations). For zero gravity-matter coupling ($G=0$), the positive-energy modes are the BEC Goldstone-Bogoliubov and the Higgs, as described in Eq. \eqref{eq_modes}, and the gravity mode $\omega=ck$. In this situation, all modes are real and therefore dynamically stable (see Fig. \ref{fig_disp} a). In the presence of gravity, however, the Bogoliubov and the gravity modes hybridize and collapse, exhibiting an imaginary part for $k-$modes below the Jeans wave vector $k_J$ that satisfy the condition $\Im(\omega)=0$, for which we obtain the equation.
\begin{equation}
\hbar ^4k_J^4 + 2 \hbar ^2 k_J^2 m^2 c_s^2 -\tilde \kappa  m^4 (c^2 + c_s^2)^2=0.
\end{equation}
Remarkably, the imaginary part of the Bogoliubov and the gravity modes have opposite signs, suggesting that the formation of dark-matter cosmological structures (triggered by the long-wavelength dynamical instability) is accompanied by the damping of space-time perturbations. In other words, the gravitational waves transfer their energy to the BEC modes so the latter can grow. Because $Re(\omega)>0$, a $I_o$-type of instability \cite{cross_93} is responsible for the formation of large structures in flat spacetimes. Also, we observe that the positive and negative Bogoliubov modes are not symmetric, i.e. $\omega_-\neq -\omega_-^*$, indicating violation of the $\mathcal{PT}-$symmetry. These features are depicted in Fig. \ref{fig_disp} b).  
\begin{figure}[t!]
\includegraphics[scale=0.47]{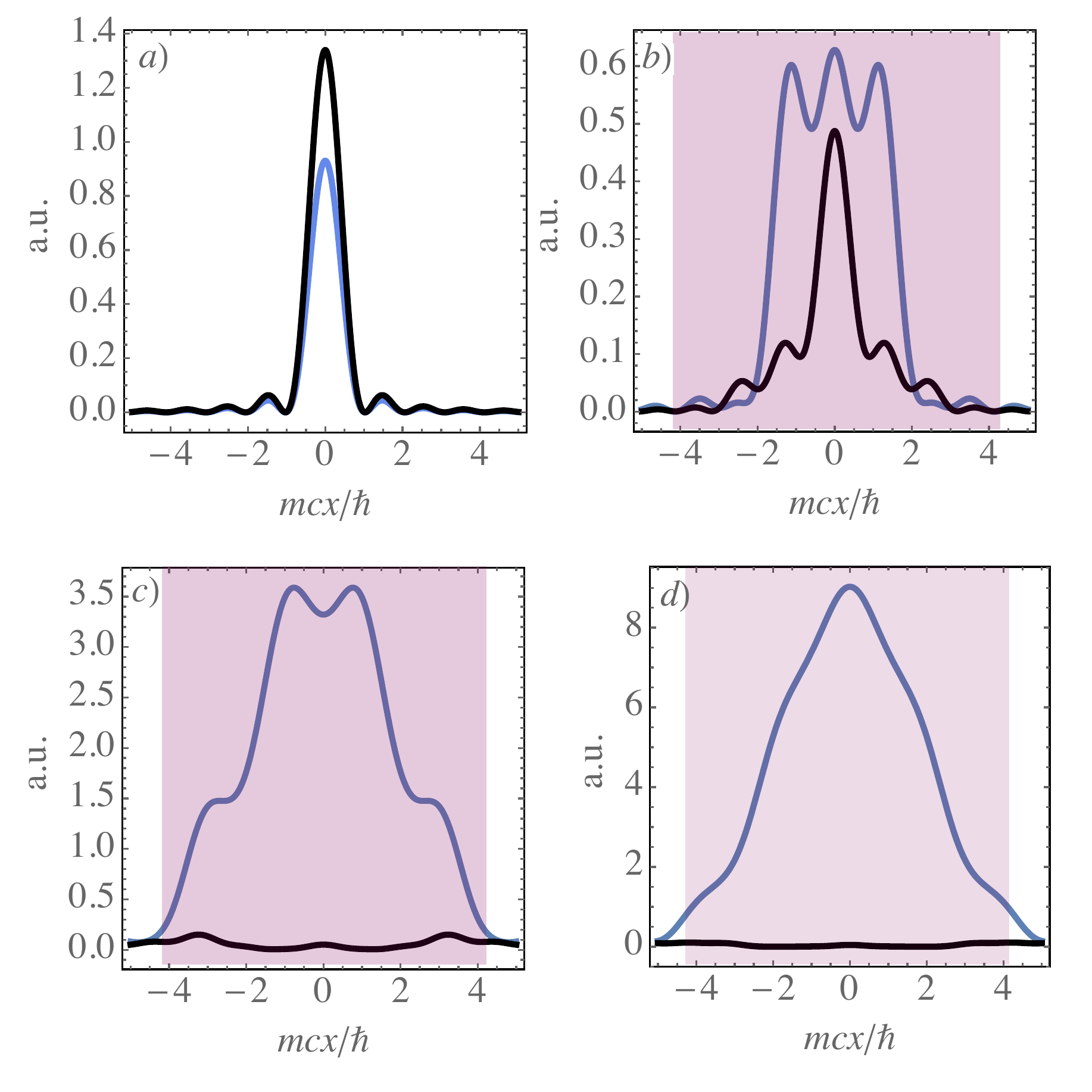}
\caption{One-dimensional illustration of the structure formation dynamics at early stages of the instability onset. Panel a) shows the initial ($t=0$) plane-wave superposition solution for the gravitational wave $\phi(x)$ (black line) and the BEC $\varphi(x)$ (lighter line). Panels b), c) and d) depict their evolution at $t=1.5 \hbar/mc^2$, $t=4.5 \hbar/mc^2$ and $t=5.5 \hbar/mc^2$, respectively. The shadowed region represent the Jeans length $\lambda_J=2\pi/k_J$. We use $c_s=c/3$. }
\label{fig_structure}
\end{figure}

In order to illustrate how the $\mathcal{PT}$-symmetry breaking affects the formation of structures, we perform one-dimensional simulations of Eqs. \eqref{eq_KG3} and \eqref{eq_E2} for the early stages of the Jeans instability. As depicted in Fig. \ref{fig_structure}, an initial linear superposition of plane gravitational and Bogoliubov waves (Fig. \ref{fig_structure} a)) lead to the formation of 1d structures in the BEC sector. Short after the onset the instability, the long wavelength structures of the BEC start to grow, leading to the formation of structures of typical size $\lambda_J=2\pi/k_J$. Simultaneously, the gravitational modes in the same wavelength modes attenuate, eventually vanishing out for longer times. We notice that our calculations are valid near the onset of the instability only, i.e for $t\ll \hbar/mc_s^2$, for which a quasi-linear approximation of Eqs. \eqref{eq_KG3} and \eqref{eq_E2} is valid. A more accurate, quantitative discussion of our results would involve taking into account saturation effects.

{\it Conclusions.}
In this work, we have studied the coupling between a gravitational wave in a Minkowski spacetime with dark matter modelled by a self-interacting complex scalar field (Bose-Einstein condensate). Considering perturbations in the spatial components of the metric only, the gravitational wave dispersion relation is analogous to that of an electromagnetic wave propagating in a charged medium characterized by a plasma frequency $\omega_p$, where the photon also acquires an effective mass. In this case, the two modes (the gravity mode and the Bogoliubov mode) are decoupled. However, when we consider perturbations in the temporal component of the metric, the gravity and the Bogoliubov modes hybridize and become dynamically unstable. Because of the local breaking of the $\mathcal{PT}$- symmetry, the modes form conjugate pairs, in such a way that there is a transfer of energy from the gravitational wave (damping) to the BEC field (growth). In short, this means that the instability mechanism triggering the formation of large dark-matter structures is accompanied by the breaking of the $U(1)\times {\cal PT}$ symmetry. Remarkably, our findings may also constitute an alternative explanation why primordial gravitational waves are quite hard to detect: they just vanish and give away their energy to the formation of large-scale structures. In a near future, our work could strongly benefit from numerical GR tools, both in weak and strong gravity scenarios, which could correctly describe the saturation at later stages due to the nonlinearity in the Klein-Gordon equation and, eventually, the effects of curvature due to the presence of massive objects.     \par
{\it Acknowledgments.} One of the authors (H. T.) acknowledges the Security of Quantum Information Group for the hospitality and for providing the working conditions, and financial support from Funda\c c\~ao para a Ci\^encia e a Tecnologia (Portugal) through grant number SFRH/BPD/110059/2015. The work of C.B. is supported by the European Research Council (ERC-2010-AdG Grant 267841).

\end{document}